\documentclass[pdflatex,sn-mathphys-num]{sn-jnl}

\usepackage{graphicx}%
\usepackage{multirow}%
\usepackage{amsmath,amssymb,amsfonts}%
\usepackage{amsthm}%
\usepackage{mathrsfs}%
\usepackage[title]{appendix}%
\usepackage{xcolor}%
\usepackage{textcomp}%
\usepackage{manyfoot}%
\usepackage{booktabs}%
\usepackage{algorithm}%
\usepackage{algorithmicx}%
\usepackage{algpseudocode}%
\usepackage{listings}%

\theoremstyle{thmstyleone}%
\theoremstyle{thmstyletwo}%

\theoremstyle{thmstylethree}%

\begin{document}

\title[Article Title]{Temporally Distinguishing Photocatalytic Dynamics}


\author[1]{\fnm{Catherine J.} \sur{Fabiano}}\email{cf37@rice.edu}

\author[1]{\fnm{Shan} \sur{Deneen}}\email{sd127@rice.edu}

\author[1]{\fnm{Yigao} \sur{Yuan}}\email{yy71@rice.edu}
\author[2]{\fnm{Luke} \sur{Kay}}\email{lk53@rice.edu}

\author*[1,2,3,4]{\fnm{Peter} \sur{Nordlander}}\email{nordland@rice.edu}
\author*[1,2,3,4]{\fnm{Naomi J.} \sur{Halas}}\email{halas@rice.edu}
\author*[1,2,3,5]{\fnm{Henry O.} \sur{Everitt}}\email{he5@rice.edu}

\affil[1]{\orgdiv{Chemistry}, \orgname{Rice University}, \orgaddress{\street{6100 Main St.}, \city{Houston}, \postcode{77005}, \state{TX}, \country{U.S.A}}}

\affil[2]{\orgdiv{Electrical and Computer Engineering}, \orgname{Rice University}, \orgaddress{\street{6100 Main St.}, \city{Houston}, \postcode{77005}, \state{TX}, \country{U.S.A}}}

\affil[3]{\orgdiv{Physics and Astronomy}, \orgname{Rice University}, \orgaddress{\street{6100 Main St.}, \city{Houston}, \postcode{77005}, \state{TX}, \country{U.S.A}}}

\affil[4]{\orgdiv{Material Science and Nanoengineering}, \orgname{Rice University}, \orgaddress{\street{6100 Main St.}, \city{Houston}, \postcode{77005}, \state{TX}, \country{U.S.A}}}

\affil[5]{\orgdiv{DEVCOM Army Research Laboratory}, \orgname{Rice University},
\orgaddress{\street{6100 Main St.}, \city{Houston}, \postcode{77005}, \state{TX}, \country{U.S.A}}}

\abstract{Light can induce both photothermal and nonthermal catalytic activity in  plasmonic nanoparticles, but the extent, timescale, and efficacy of these two mechanisms remain unresolved. Here we introduce pump-pump photocatalysis, an adaptation of ultrafast excitation correlation spectroscopy in which incident laser pulses are split into two spatially and energetically equivalent  pulses separated by a variable time delay. For a given reaction and catalyst, time-sensitive nonlinear enhancements in photocatalytic activity may be distinguished from time-insensitive responses as a function of time delay, pulse power, excitation wavelength, applied temperature, reactant pressure, and pulse asymmetry. In this way, the timescales of photothermal and nonthermal activity of any photocatalyst may be monitored on sub picosecond to nanosecond timescales, and the conditions for optimal chemical reactivity may be discovered. 
Here we report ultrafast measurements of the ammonia decomposition reaction using a Cu-Ru antenna-reactor photocatalyst. Photothermal contributions exhibit little sensitivity to pulse delay, while nonthermal contributions are most apparent at low excitation intensity, moderate temperatures, and sub-nanosecond timescales. Here, nonlinear, nonthermal mechanisms enhance H$_2$ production by a factor up to eleven compared to when pulses overlap. This technique may be used to provide unprecedented \textit{in operando} diagnostics and control of any photocatalyst for any chemical reaction. Most importantly, these measurements allow us to ascertain the optimal distribution of light to maximize photocatalytic activity.}

\keywords{plasmonic photocatalysis, ultrafast spectroscopy, Keyword3, Keyword4}



\maketitle

\section{Introduction}\label{sec1}
Over the last two decades, a growing body of work has explored the illumination of tailored photocatalysts to demonstrate unprecedented chemical reactivity, selectivity, and control beyond what is possible with traditional heterogeneous catalysis.\cite{yigaoScience,christopher2012singular, linic2015photochemical}. Questions have arisen as to the mechanisms of these photo-induced enhancements. Such mechanisms include photothermal and nonthermal mechanisms, such as those mediated by hot carrier distributions, providing new pathways for enhancing reaction rates, reducing activation barriers, improving product selectivity, or altering the rate determining step \cite{zhang2017product,linanScience, christopher2012singular,linic2013catalytic}. A variety of experimental and theoretical techniques have been used to confirm or challenge the claims associated with these nonthermal mechanisms, typically characterized by the assertion that local temperature, when adequately accounted, can explain these favorable results\cite{li2020synergy,baffou2020simple,antennareactorPNAS,yigaoScience}.  Unfortunately, adequate techniques for measuring instantaneous local temperature of nanometer-sized photocatalysts $\textit{in operando}$ are not yet available.

Of particular interest for this topic have been plasmonic metal nanoparticles (NPs), used to transform light into chemical energy through photothermal heating and the generation of hot carriers (HCs)\cite{brongersma}. 
Despite the success of plasmonic photocatalysis, it has proven challenging to distinguish between photothermal and nonthermal contributions experimentally because of the synergy between these processes\cite{li2020synergy,zhang2018plasmon,li2019confirming, geng2023achieving}. The dramatic reduction of activation barriers induced by hot carriers still requires modest thermal excitations for reactions to proceed \cite{linanScience}. Wavelength-dependent measurements of rates and selectivities have provided a clear proof of hot carrier generated chemistry \cite{liu2025quantifying, yigaoScience,christopher2011visible} since the optimal selectivities and maximum reaction rates typically occur at wavelengths significantly shifted away from the absorption maximum of the plasmonic photocatalyst \cite{yuan2024steam}.
 Previous studies have begun to distinguish between photothermal and nonthermal contributions by focusing on measuring or calculating the surface temperature\cite{nanotherm, hu2018quantifying,baffou2010thermoplasmonics} and by quantifying the plasmonic resonance response through power and wavelength dependence \cite{christopher2012singular, linanScience}. However, none of these techniques incorporate the most distinguishing factor between photothermal and nonthermal effects - their timescales. Nonthermal effects occur on much faster timescales than photothermal effects, as the former involve the ultrafast excitation and relaxation of carriers over hundreds of femtoseconds to picoseconds, while the latter involve the slower heating of the lattice over hundreds of picoseconds to nanoseconds\cite{brongersma}.

\begin{figure} 
	\centering
	\includegraphics[width=1\textwidth]{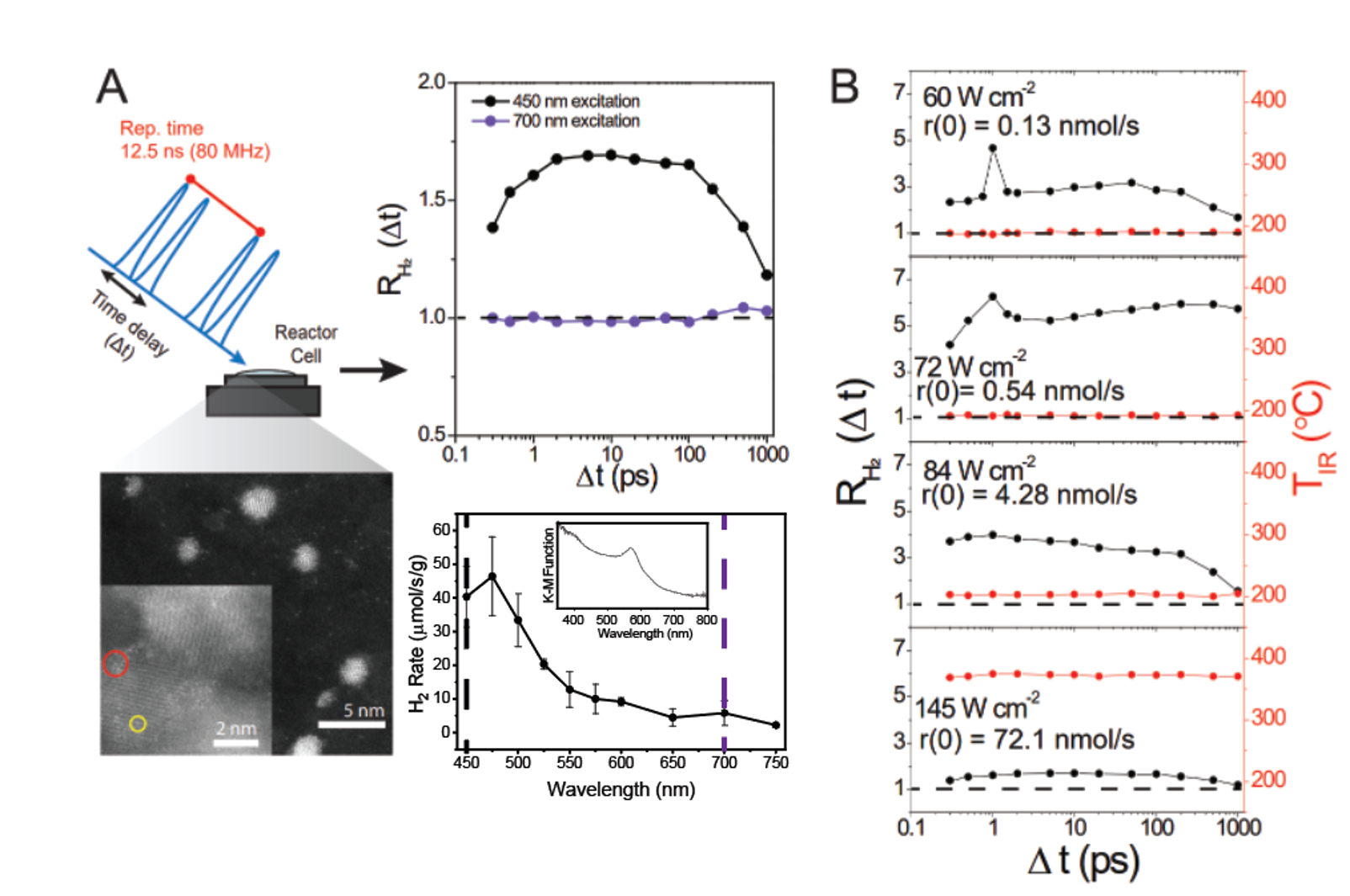} 

	\caption{\textbf{Pump-Pump Experimental Technique.}
(\textbf{A}) Illustration of the pump-pump technique and transmission electron microscope image of the Cu-Ru AR photocatalyst, with inset indicating the location of Ru atoms (yellow circle) and clusters (red circle). Product ratio $R_{H_2}(\Delta t)$ for NH$_3$ decomposition as a function of pump pulse delay $\Delta t$ for two wavelengths, one near-resonant (450 nm) and one not resonant (700 nm) with the plasmon resonance of the photocatalyst with 145 $W/cm^2$. Measured action and diffuse reflectance (inset) spectra of the Cu-Ru AR photocatalyst, from \cite{yigaoScience}.  (\textbf{B}) Rate enhancement ratio $R_{H_2}(\Delta t)$ for four pulse intensities, illustrating that the nonlinear catalytic response depends strongly on pulse delay but the temperature does not. The response becomes more linear, the production rate increases ($r(0) \times R_{H_2}(\Delta t))$, and the measured temperature $T_{IR}$ increases with increasing excitation intensity.} 

	\label{fig:A} 
\end{figure}

Here we introduce a new time-resolved technique to distinguish between photothermal and nonthermal processes.
Pump-pump photocatalysis (P3C) is analogous to pump-probe and excitation correlation spectroscopy in that each pulse from the laser is split into two pulses and the temporal delay between them is adjusted. But unlike pump-probe spectroscopy, where the second "probe" pulse is much weaker and may be a different wavelength than the first "pump" pulse, both pulses in pump-pump excitation correlation spectroscopy can be spatially and energetically identical. In our experiments, these two pulses illuminate the heterogeneous photocatalyst in a reactor where the photocatalytc reaction will occur. 
We then measure the total product formation for a given temporal delay between pulses. If the system responds only to the total number of photons, regardless of when they arrive, there is no ”correlation” between the pulses, resulting in no changes in product formation. This is a linear response (Fig.\ref{fig:A}A). By contrast, if it matters when the photons arrive, such as when the first pulse excites the system into a more active (or more dormant) state, then when the delayed second pulse arrives it is more (or less) effective than when the two pulses arrive simultaneously. In this case, the system has a nonlinear response to the pulses: the excitation of the system by the second pulse is ”correlated” with the first pulse over the timescale the catalyst exhibits a nonlinear response (Fig.\ref{fig:A}).  

To differentiate between linear and nonlinear contributions, we plot our results as a product rate ratio $R_{product}(t)$ defined as
\begin{equation}\label{eq:rateratio}
    R_{product}(\Delta t) = \frac{r(\Delta t)}{r(0)},
\end{equation}
where $r(t)$ is the measured rate for product generation (\textit{e.g.} nmol/sec), with the first pulse occurring at time $t = 0$ and the second pulse occurring at time $t=\Delta t$. $R_{product}(0)=1$ is the rate ratio when the two pulses overlap. When the delayed second pulse produces $R_{product}(\Delta t)=1$, the catalyst is in the linear regime. When $R_{product}(\Delta t)>1$, delaying the second pulse enhances production, while $R_{product}(\Delta t)<1$ indicates reduced production. In either case, the amount $R_{product}(\Delta t)$ deviating from 1 indicates the degree of temporal nonlinearity exhibited by the photocatalyst. 

To illustrate the insight that can be obtained from P3C, we examine the decomposition of NH$_3$ into H$_2$ and N$_2$ using a plasmonic photocatalyst illuminated by a wavelength-tunable laser producing $\sim$~150 fs pulses with a repetition period of 12.5 ns.  
In P3C, the laser pulses are split into two spatially and energetically equivalent pulses separated by a variable time delay. As the photocatalyst is illuminated by this train of pulse pairs, the formation of the product (H$_2$) is monitored by gas chromatography as a function of pulse delay $\Delta t$ for total incident laser excitation intensity $I_{ex} (= P_{ex}/A_{ex}$ with measured laser power $P_{ex}$ and laser spot area $A_{ex}$), external heating or cooling applied to the catalyst $T_{ex}$, partial pressure of the reactant $p_{NH_3}$, and pulse asymmetry $f_1$. For our experiments, $P_{ex}$ ranged from 35 to 120 mW, while the area of the laser spot was typically 0.072 mm$^2$. Although the average temperature of the catalyst surface $T_{IR}$ is measured by an infrared camera, the instantaneous temperature of the catalyst $T(t)$ changes dynamically as the carriers and lattice are heated and cooled by the laser pulses.

We chose the NH$_3$ decomposition reaction because it has been well studied and has few side reactions\cite{bradford1997kinetics, ertl1980mechanism,li2019light,yigaoScience,YigaoACSC2026}. 
To decompose NH$_3$, a plasmonic Cu-Ru surface alloy antenna-reactor(AR)  was utilized, where the AR consists of a 7.5 nm diameter Cu nanoparticle with Ru reactor sites on the surface in a 97.5-2.5\% atomic ratio\cite{linanScience,yigaoScience,coprecip1,coprecip2}. 
Antenna-reactor complexes combine the plasmonic properties of metal NPs (antenna) with active reactor sites of transition metals (reactor)\cite{antennareactorPNAS}.  Recent studies have shown Cu-Ru and Cu-Fe ARs as efficient catalysts for NH$_3$ decomposition, as Ru\cite{linanScience} and Fe\cite{yigaoScience} are ideal binding sites predicted by Sabatier's principle\cite{sabprinciple}.

An example of a linear or nonlinear response to pulse delay $\Delta t$ may be observed when the excitation energy is far from ($\lambda_{ex}=700$ nm) or near to ($\lambda_{ex}=450$ nm) the optimal excitation wavelength ($\lambda_{opt}=475$ nm) for this reaction using the Cu-Ru AR \cite{yigaoScience} and a combined high pump intensity of $I_{ex} = 145$ W/cm$^2$ (Fig.\ref{fig:A}A). For nonresonant excitation the H$_2$ production rate remains a constant $R_{H_2}(\Delta t)=1$ regardless of the pulse time delay - a linear result characteristic of a photothermal mechanism.  In contrast, upon near-resonant excitation the H$_2$ production rate increases ($R_{H_2}(\Delta t)>1$) by an amount that depends on pulse delay - a nonlinear result. Because the temperatures are nearly identical in the nonresonant and near-resonant cases ($T_{IR} = 375^\circ$C vs $370^\circ$C, respectively), this nonlinear enhancement indicates a correlation between the first and second pulse caused by a nonthermal mechanism excited in the Cu-Ru AR. Indeed, P3C reveals that near-resonant excitation of the photocatalyst produces a nonlinear optical response that strengthens over the first picosecond and decreases after 100 ps, during which time it produces up to 70\% more product ($R_{H_2}(1-100~ps)=1.7$) than if the pulses had not been delayed. The mechanisms responsible for this nonlinear response will be discussed shortly, but for now it is sufficient to recognize that delaying the second pulse improved H$_2$ production using the same number of photons and maintaining the same temperature. Thus, the first pulse excited the catalyst in a nonlinear manner, making the second pulse more effective when delayed by $t=\Delta t$ than when arriving with the first pulse at $t=0$. 
With only one exception, we found that $R_{H_2}(\Delta t)\ge~1$ for every experimental condition.  The exception occurred when the Ru was removed from the Cu-Ru AR, producing a time-independent $R_{H_2}(\Delta t)=0.5$, suggesting that no reactants remained for the second pulse (see SM). This observation is consistent with Sabatier's principle that Ru supports NH$_3$ binding more efficiently than Cu\cite{sabprinciple}.  

Staying on resonance, we now investigate how the nonlinear catalytic activity depends on excitation intensity. As the laser intensity $I_{ex}$ increases from 60 to 145 W/cm$^2$, Fig.\ref{fig:A}B reveals that the overall production of H$_2$ increases as expected (see $r(0)$). However, $R_{H_2}(\Delta t)$ decreases from large enhancements (2x to 6x) toward none (1x), the temporal response flattens, and the temperature rises, indicating a transition of the decomposition mechanism from a nonlinear to a linear process.  The nonlinear temporal effects ($R_{H_2}(\Delta t)\ge~1$) are strongest at lower powers and temperatures and become increasingly linear at higher powers and temperatures, suggesting that nonthermal mechanisms are most clearly observed in the lower power regime while photothermal catalysis dominates in the higher power regime.  We discovered that at the lowest power density ($I_{ex} = 60~$W/cm$^2$) the maximum H$_2$ production rate occurs narrowly near $\Delta$t = 1 ps (Fig.\ref{fig:A}B top), indicating an additional mechanism beyond the broader, slower nonlinear response observed in Fig.\ref{fig:A}A \cite{Cat1}. Also, the character of the broader nonlinear response changes with pump intensity, with maximum production occurring for delays near or beyond 100 ps for the lower intensities, then shifting to earlier times for higher intensities.

\begin{figure} 
	\centering
	\includegraphics[width=1.1\textwidth]{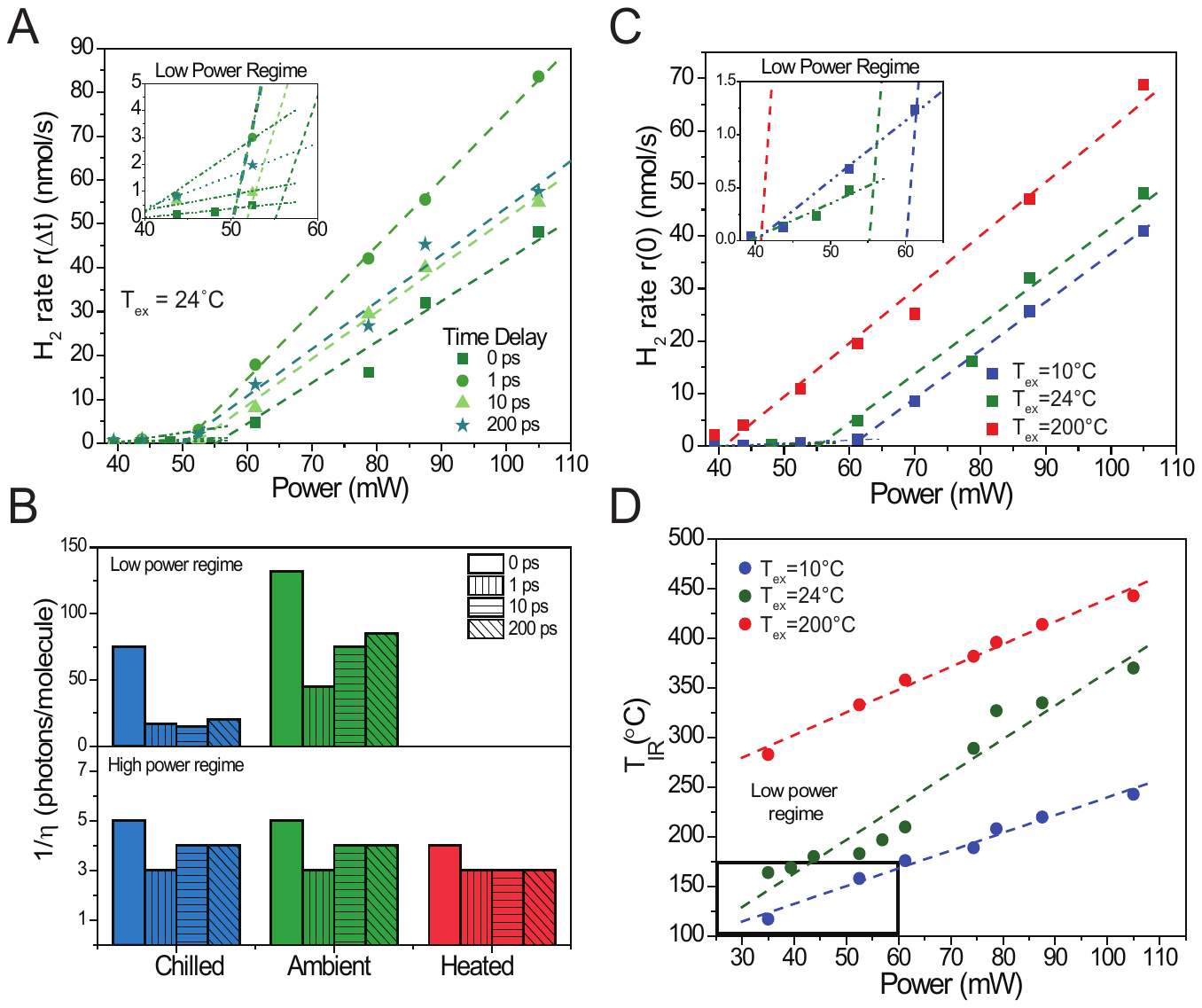} 

	\caption{\textbf{Temporal and Temperature Relationship.}
	(\textbf{A})The H$_2$ formation rate $r(t)$ as a function of $P_{ex}$ for $\lambda_{ex}=450$ nm at four pump-pump delays: $\Delta t =$ 0, 1, 10, and 200 ps. (\textbf{B}) Comparison of H$_2$ efficiency ($1/\eta$ photons/molecule) in the low and high power regime for these same delays. (\textbf{C}) The H$_2$ formation rate $r(t)$ as a function of laser power $P_{ex}$ for $\lambda_{ex}=450$ nm at $\Delta t= 0$ ps for three reaction chamber conditions: chilled ($T_{ex}=10^\circ$C, blue), ambient ($T_{ex}=$ room temperature, green) and heated ($T_{ex}=200^\circ$C, red). (\textbf{D}) The measured temperature $T_{IR}$ as a function of $P_{ex}$ for these same conditions, illustrating the low power region where the average surface temperature is below the 175$^\circ$C required for associative desorption of H$_2$.}
	\label{fig:C} 
\end{figure}

To understand this evolution, P3C can measure how rate $r(\Delta t)$ depends on excitation power $P_{ex}$ for a given pulse delay $\Delta t$.  For the NH$_3$ decomposition reaction, we observe $r(\Delta t)$ increases linearly with increasing $P_{ex}$ for each pulse delay, including $\Delta t = 0$. Interestingly, Fig.\ref{fig:C}A reveals two linear regimes, one at low powers and one at high powers, each characterized by a linear increase in rate with power. In each regime we find a minimum "threshold" power $P_{th}$ is required to produce H$_2$, above which $r(t)$ increases linearly with increasing $P_{ex}$ as
\begin{equation}\label{eq:NLeq}
    r(t)=\eta (P_{ex} - P_{th}),
\end{equation}
where $\eta_i$ is the efficiency of additional photons to produce H$_2$ (e.g. nmol/W or molecules/photon) when $P_{ex} > P_{th}^i$ for the low or high power regimes ($i=L,H$, respectively). Three important insights immediately arise. First, a large number of "precursor" photons are required to prepare (\textit{i.e.} heat) the catalyst before the reaction produces any products. Second, above this threshold the number of "converting" photons required to produce additional products is a constant $1/\eta$ photons per molecule. Third, the $linear$ increase in $r(\Delta t)$ with $P_{ex}$ occurs even in the low power regime where these same rates exhibit a $nonlinear$ dependence on pulse delay.  Summarizing, $P_{th}$ measures how much photothermal heating is required for the reaction to begin, catalysts or regimes with larger $\eta$ need fewer photons to produce additional products, and both $P_{th}^i$ and $\eta_i$ depend on pulse delay $\Delta t$.

Exploring now the two power regimes, we find that both $P_{th}$ and $\eta$ are smaller in the low power regime than the high power regime (see Table 1 and Fig.\ref{fig:C}B). Thus, the low power regime has a lower onset power ($i.e.$ temperature) but produces fewer products above threshold.  Although the values of $P_{th}^L$ are relatively insensitive to $\Delta t$ in the low power regime (see Table 1), $\eta_L$ depends more sensitively on $\Delta t$, with $\Delta t = 0$ having the lowest efficiency and $\Delta t = 1$ ps having the highest (Fig.\ref{fig:C}B). The higher values of $P_{th}^H$ and $\eta_H$ in the high power regime suggest that once photothermal heating raises the catalyst above a certain temperature $T_{th}^H$, decomposition accelerates and operates at a much more efficient rate ($\eta_H > 10~\eta_L$). In this photothermal regime, production is linear in both power and time. 
The values for $P_{th}^H$ and $\eta_H$ in the high power regime depended much less on $\Delta t$, as expected from Fig.\ref{fig:A}B. Note that the average temperature $T_{IR}$ of the catalyst at $P_{th}^H$ is 168 - 185$^\circ$C for each delay, while at $P_{th}^L$, $T_{IR}$ spans 108 - 121$^\circ$C.


 \begin{table} 
	\centering
	\caption{\textbf{Power-dependent Photocatalysis for Ambient Conditions.}
		}
    \label{tab:example} 

	\begin{tabular}{c|ccc|ccc} 
		\\
	\hline
		& Low~Power~Regime &${} $ &${} $ & High~Power~Regime & ${}$ & ${}$ \\
        Delay & $1/\eta_L$ & $P_{th}^L$ & $T_{th}^L$ & $1/\eta_H$ & $P_{th}^H$& $T_{th}^H$ \\
		(ps) & (photons/molecule) & (mW) & ($^\circ$C) & (photons/molecule) & (mW) & ($^\circ$C)\\
		\hline
		0  & 132 & 44 & 121 & 5 & 63 & 185\\
	    1 & 45 & 40 & 108 & 3 & 58 & 168\\
        10 & 75 & 40 & 108 & 4 & 59 & 172\\
        200 & 85 & 40 & 108 & 4 & 48 & 168\\
		\hline
	\end{tabular}
\end{table}

Since temperature is such a critical parameter for all catalytic reactions, including photocatalytic ones, we now use P3C to explore the relationship between laser intensity and externally applied heating or cooling. We start with intensity-dependent measurements for $\Delta t = 0$ ps after either externally chilling ($T_{ex} = 10^\circ$C) the catalyst, heating it ($T_{ex} = 200^\circ$C), or leaving it at ambient temperature ($T_{ex} =24^\circ$C). Temperature control was achieved using an external chilling line and a heating element within the reactor for cooling and heating, respectively (SM). 
Fig.\ref{fig:C}C reveals that both low and high power regimes are present at chilled and ambient temperatures, but only the high power regime is present in the heated case (Table 1). The latter is synergy: external heating and light combine to improve the performance of the catalyst beyond what is possible with either alone \cite{li2019confirming,li2020synergy,zhang2018plasmon}. 

An equally important insight may be inferred by exploring the low power regime for the chilled and ambient cases. Fig.\ref{fig:C}D plots how the measured temperature $T_{IR}$ increases with $P_{ex}$.  Notice that the low power regime occurs when $108^\circ$C $< T_{IR} < 175^\circ$C, while the high power regime begins when $T_{IR} > 175^\circ$C.  It is known that dissociative adsorption of NH$_3$ on a crystalline Ru (0001) surface begins near $T_{da}$ = 127$^\circ$C (400 K), while thermal associative desorption of H$_2$ from Ru begins at $T_{ad}$ = 177$^\circ$C (450 K) \cite{mortensen2000dynamics, ZUPANC2002501}. Although these onset temperatures $T_{da}$ and $T_{ad}$ will be different for nanostructured Cu-Ru ARs, the P3C data reveal that the low power regime is likely characterized by photothermal heating sufficient for dissociative adsorption of NH$_3$, while the transition from the low to high power regimes likely occurs as the catalyst surface reaches the temperature for associative desorption of H$_2$ by heating and/or by illumination. 

If so, this would explain why the efficiency of H$_2$ production $\eta_L$ is so much lower than $\eta_H$: hydrogen produced by NH$_3$ decomposition struggles to desorb in the low power regime. Notice that the low power regime extends to higher laser powers in the chilled case than in the ambient case, since it takes more photothermal heating to reach the high power regime there. 
Notice also that the highest threshold $P_{th}^L$ and lowest efficiency $\eta_L$ occur for $\Delta t =$ 0 (Fig.\ref{fig:C}A, Table 1), confirmation that H$_2$ production increases when the second pulse is delayed. This suggests that the nonlinear temporal behavior observed in this regime is caused by a mechanism that induces H$_2$ desorption at temperatures below $T_{ad}$.
Also notice that once the high power photothermal regime is entered, the efficiency of the reaction is the same ($1/\eta_H = 3-5$ photons/molecule) regardless of the source of heating. Interestingly, P$_{th}^H$ for the heated case is roughly the same as P$_{th}^L$ for the low power regime in the ambient and chilled cases, indicating that the heated catalyst is always in the high power regime where the rate determining step (RDS) is NH$_3$ dissociative adsorption.  

\begin{figure} 
	\centering
	\includegraphics[width=1\textwidth]{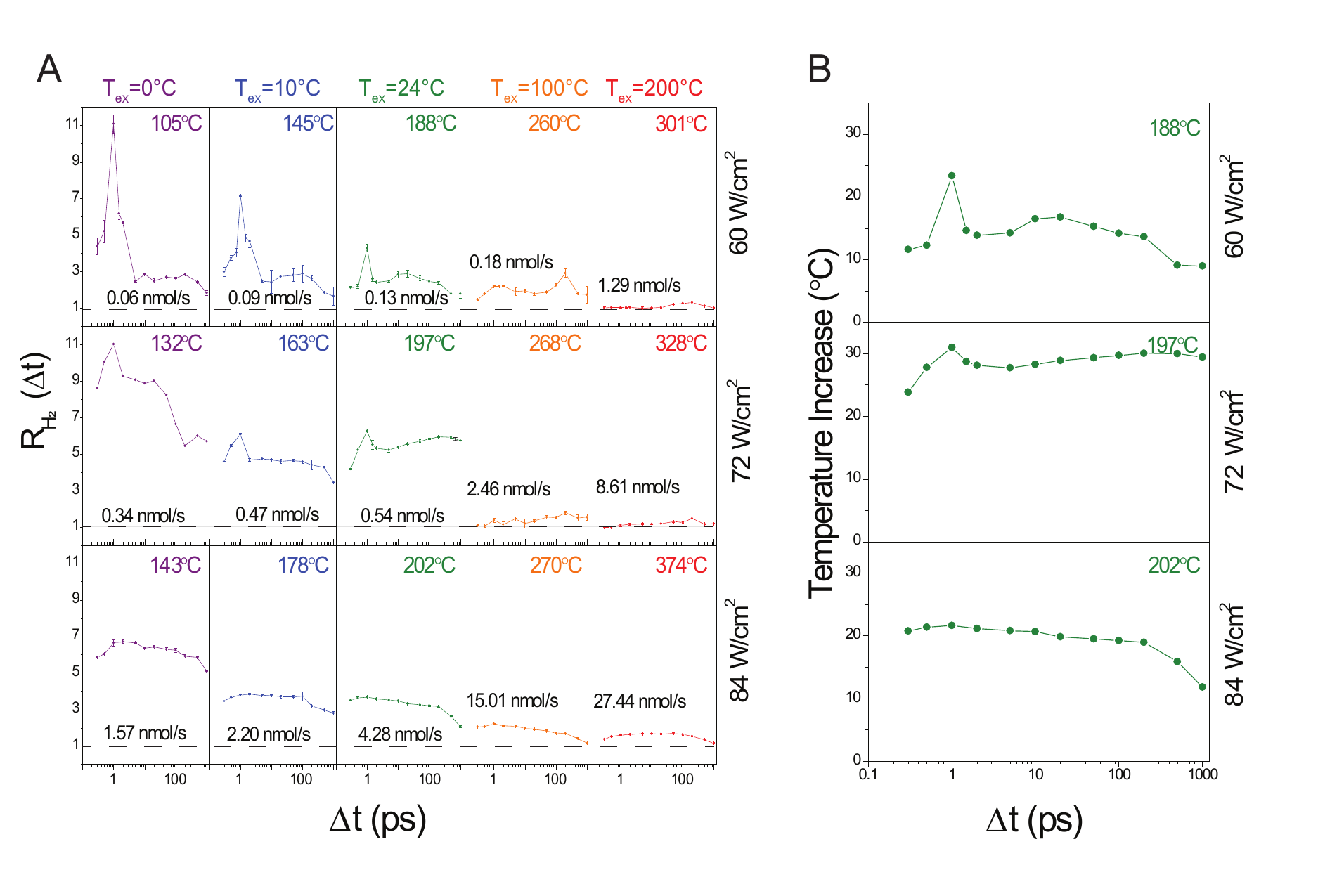} 

	\caption{\textbf{Time Evolution as a Function of $T_{ex}$.}
	(\textbf{A}) H$_2$ rate ratio $R_{H_2}(\Delta t)$ as a function of external cooling and heating of the reaction chamber for three power densities ($I_{ex}$ = 60, 72, and 84 W/cm$^2$). The black dashed line represents $R_{H_2}(\Delta t)$ = 1 with the corresponding H$_2$ formation rate $r(0)$ listed. (\textbf{B}) Estimate of the required rise in transient temperature $T(\Delta t)$ above the measured $T_{IR}$ (green number) to explain the observed delay-dependent nonlinear production rates, assuming a constant activation energy of 1.2 eV.} 
	\label{fig:B} 
\end{figure}

The connection between temporal behavior, laser power, and external heating/cooling is further clarified in Fig.\ref{fig:B}A, where the rate enhancement ratio $R_{H_2}(\Delta t)$ is plotted for all pulse delays with $T_{ex}$ spanning 0 - 200$^\circ$C and $I_{ex}$ spanning the low (60 W/cm$^2$) to high power regimes (84 W/cm$^2$). Reading from lowest to highest temperature (left to right), we see the transition from the now familiar nonlinear, nonthermal behavior to linear, photothermal behavior as $R_{H_2}(\Delta t) \rightarrow 1$. Focusing next on the low power regime with $I_{ex} = 60~$W/cm$^2$ (top row), in addition to the broad nonlinear response with $R>$ 1 we discover a startling new feature at 1 ps: H$_2$ production enhancement over the $r(0)$ rate that rises from $R_{H_2}$(1 ps) = 4 under ambient conditions to an astonishing $R_{H_2}$(1~ps) = 11 for the cooled catalyst \cite{Cat1}. This narrow peak shrinks rapidly as $T_{ex}$ or $I_{ex}$ increase.

The P3C data may be used to provide an understanding of this behavior. 
For example, if we assume a purely thermal model with a constant activation energy for H$_2$ desorption from the Cu-Ru AR, then increases in $r(\Delta t)$ above $r(0)$ must be caused by changes in transient temperature $T(\Delta t)$ above $T_{IR}$. To explore that hypothesis, Fig.\ref{fig:B}B plots the deduced catalyst temperature increase for the 60, 72, and 84 W/cm$^2$ cases assuming a constant H$_2$ desorption activation energy of 1.2 eV \cite{linanScience}. The time-dependent rise in $T(\Delta t) - T_{IR}$ needed to explain the nonlinear rise in reaction rate caused by the second pulse is between $10-30^\circ$C. Although this is a small increase in transient temperature, it is approximately ten times larger than reported in recent estimates of photoinduced thermal transients in a nearly identical system \cite{YigaoACSC2026}. 
Moreover, even this modest amount of additional heating cannot explain the data, because as noted above, the nonlinear temporal behavior is only observed with near-resonant excitation, not when excitation is far from resonance for any delay $\Delta t$, even though both produce the same $T_{IR}$.  Thus, the nonlinear increase in reaction rate cannot be purely thermal, ruling out that hypothesis. 

A hypothesis that does explain the observed behavior stipulates that applied heating, whether dark or photothermal, cracks adsorbing NH$_3$ when $T_{IR} > 100^\circ$ C \cite{mortensen2000dynamics}. In this low power, low temperature regime, the reaction rates are low because the molecular fragments and products NH$_x$ and H$_2$ remain adsorbed on the surface of the Ru reactors, poisoning the surface and suppressing new reactions. However, when excited on resonance, photogenerated hot electrons act to desorb H$_2$ in a way heating cannot, producing reaction rates many times higher than when the two pulses coincide. Nevertheless, this nonlinear mechanism is inefficient, requiring $1/\eta^L = 45-132$ photons to desorb one $H_2$ molecule. When additional dark or photothermal heating raises the temperature toward 200$^\circ$ C \cite{ZUPANC2002501}, H$_2$ is thermally desorbed, the surface is no longer poisoned, nonthermal effects are no longer needed, and the high temperature regime is entered where it only takes an additional $1/\eta^H = 3-5$ photons to liberate a hydrogen molecule.

The most dramatic of the nonthermal effects is observed near 1 ps, when the second pulse photogenerates new hot electrons precisely when the hot electrons generated by the first pulse reach their optimal temperature. These dramatic effects are addressed in a separate manuscript \cite{Cat1}. In addition to this highly nonlinear feature, only observed at low powers and temperatures near 1 ps, the data reveal a broad nonlinear enhancement of $R(\Delta t)=2-6$ over most subnanosecond time scales, including a second enhanced nonlinearity near 200 ps when external heating is combined with low laser power. Since the lattice is cooling 200 ps after the first pulse, this feature cannot be explained by changes in temperature or activation energy.  Instead, this feature may represent diffusion of adsorbed NH$_3$ from the Cu antenna to the cleared Ru reactor following the first pulse.  This feature is not seen at lower temperatures because diffusion is slower, nor is it seen for higher powers because the reaction is so much faster and more efficient.

\begin{figure} 
	\centering
	\includegraphics[width=1\textwidth]{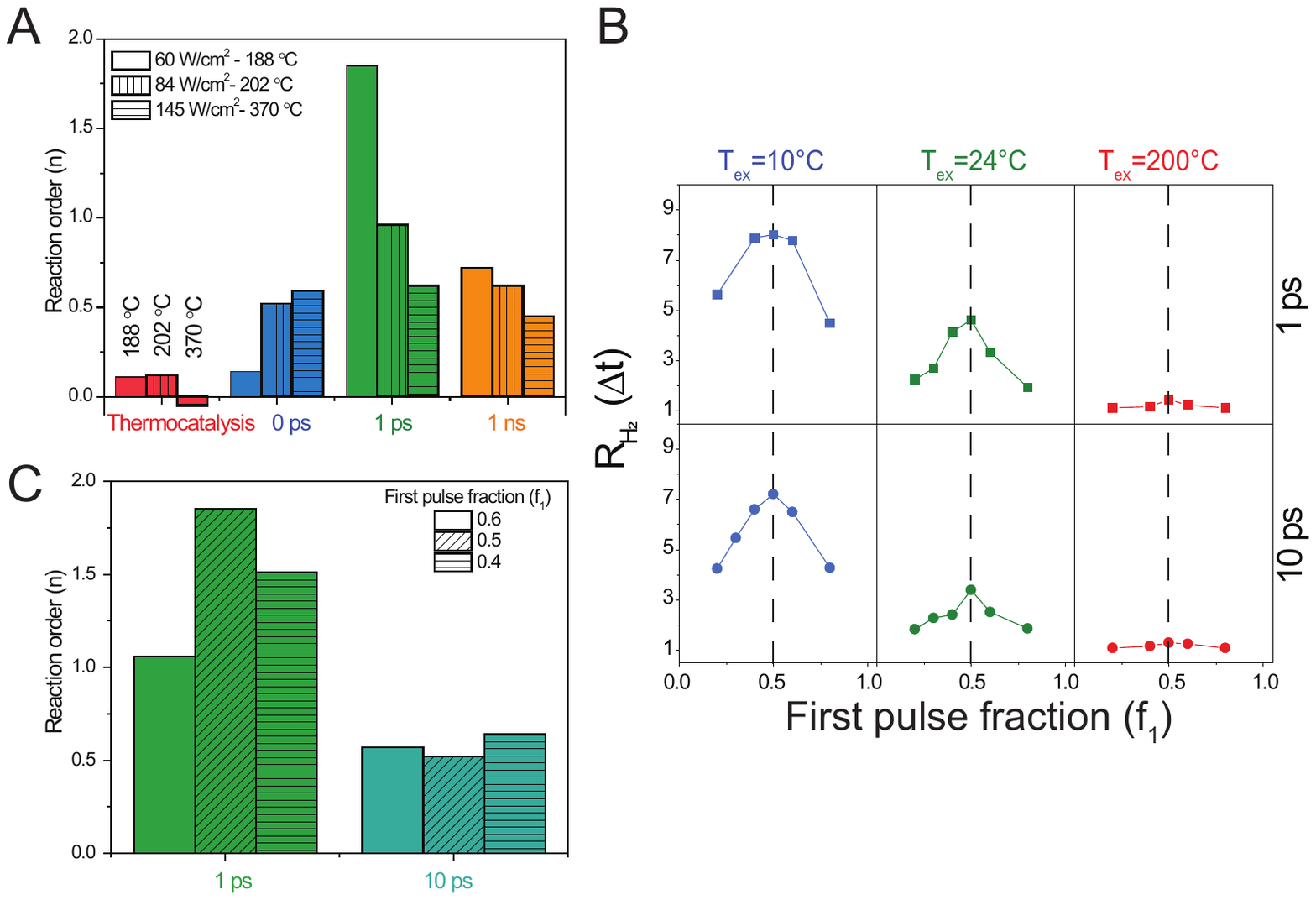} 

	\caption{\textbf{Reaction order}
	(\textbf{A}) Comparison of reaction order $n$ as a function of power density $I_{ex}$ and time delay $\Delta t$. (\textbf{B}) H$_2$ rate ratio $R_{H_2}(\Delta t)$ as a function of pulse asymmetry $f_1$ when the reaction chamber is externally cooled (blue), at ambient temperature (green), and heated (red) for a power density of $I_{ex}$ = 60 $W/cm^2$ at two delays ($\Delta$t = 1 and 10 ps). (\textbf{C}) Comparison of reaction order (n) as a function $f_1$ for $\Delta$t = 1 and 10 ps.}

	\label{fig:D} 

\end{figure}

Two final capabilities of P3C will be used to confirm these conjectures.  The first is its ability to measure the production rate as a function of reactant partial pressure to ascertain how reaction order $n$ depends on pulse delay and intensity. Such measurements reveal the RDS for each $\Delta t$. For NH$_3$ decomposition, two RDSs are known: (i) N-H bond breaking ($n=1$) and (ii) N$_2$ desorption ($n=0$)\cite{tsai1987steady}. 
Using a mixture of He, Ar, and NH$_3$ to maintain constant temperature and total pressure, the measured reaction rate reveals a zeroth-order dependence on NH$_3$ partial pressure $p_{NH_3}$ for dark thermocatalytic reactions but a roughly first-order ($0.5 < n < 1.0$) dependence for almost every delay $\Delta t$ and intensity $I_{ex}$ (Fig.\ref{fig:D}A). Thus, as inferred above, the RDS for most of our measurements at most times for most powers, is NH$_3$ dissociative adsorption.   

Interestingly, for the case where $R_{H_2}$(1~ps) = 4 in the ambient low power regime ($I_{ex} = 60$~W/cm$^2$), a second-order reaction is observed, indicating that the photocatalyst is profoundly starved of reactant. 
Clearly a different mechanism is at work here. 
As described elsewhere \cite{Cat1}, the most plausible mechanism is hot carrier-induced breaking of the N-H bond followed by associative desorption of H$_2$. Indeed, this process is so much more efficient than photothermal desorption that the catalyst is quickly rid of adsorbed H$_2$ and therefore starved of reactant NH$_3$.
Thus, delaying the second pulse plays a more pivotal role in the low power regime when the temperature of the surface is lower and a nonlinear, nonthermal mechanism is required for H$_2$ desorption.

Finally, to confirm that the second pulse facilitates photodesorption in the low power regime, P3C allows us to unbalance the two pulse intensities. This unbalancing allows us to explore whether the reaction favors a second pulse that is stronger or weaker than the first for a given $\Delta t$ and total intensity $I_{ex}$. We define $f_1$ as the first pulse fraction: $f_1$ = 0.2 means the first pulse contains 20\% of $I_{ex}$ and the second pulse contains 80\%. For all conditions considered here (Fig.\ref{fig:D}B), the greatest rate enhancement $R_{H_2}(\Delta t)$ occurs when the pulses are balanced ($f_1 =$ 0.5). To discover whether a reaction favors a stronger second pulse, we unbalance the pulses and compare $R_{H_2}(\Delta t)$ for $f_1$ = 0.4 with $f_1$ = 0.6 and $f_1$ = 0.3 with $f_1$ = 0.7. If the pairs give the same $R_{H_2}(\Delta t)$, then the ordering of the unbalanced pulses does not matter, and a plot of $R_{H_2}(\Delta t)$ vs $f_1$ is symmetrical about $f_1=0.5$. We see this behavior, for example, when $\Delta t = 10$ ps for all three values of $T_{ex}$ (Fig.\ref{fig:D}B).  

However, if the ordering of the unbalanced pulses matters, the plot will be asymmetric and a reaction that favors a stronger second pulse will be skewed in favor of $f_1 <$ 0.5. This kind of unbalance is observed at $\Delta t$ = 1 ps (chilled or ambient) in the low power regime ($I_{ex} = 60~W/cm^2$). 
This asymmetry confirms that the first pulse decomposes the NH$_3$ while the stronger second pulse induces the photoexcited catalyst to desorb the H$_2$ product for temperatures below $T_{ad}$.
Additional confirmation comes from repeating the partial pressure experiment but with $f_1 =$ 0.4, 0.5, and 0.6 to compare reaction order $n$ for $\Delta t =$ 1 and 10 ps (Fig.\ref{fig:D}C). When $\Delta$t = 10 ps, a first-order reaction is observed for all values of $f_1$, but when $\Delta t$ = 1 ps, 
the reaction order depends on $f_1$. Here, the second-order reaction originally seen for $f_1 =$ 0.5 is also observed for $f_1 =$ 0.4, while only a first-order reaction is observed for $f_1 =$ 0.6. 
This indicates that the reaction is more starved for reactants when the first pulse is weaker, presumably because the stronger second pulse is more effective at desorbing products bound to the surface through a nonthermal photodesorption mechanism when the catalyst is in a nonequilibrium state.


In conclusion, we have introduced pump-pump photocatalysis (P3C), a variation of excitation correlation spectroscopy, to measure and analyze linear and nonlinear photothermal and nonthermal processes $\textit{in operando}$. We use the well-known NH$_3$ decomposition reaction on a Cu-Ru antenna-reactor photocatalyst activated by a wavelength-tunable pulsed laser. A low power, low H$_2$ production regime 
is characterized by nonthermal desorption of H$_2$ in a manner that varies nonlinearly with pulse delay. When the second pulse is delayed 1-100 ps, we observe an H$_2$ production enhancement $2-11$ times greater than the undelayed 0 ps case. By contrast, a high-power regime is observed in which the catalyst is hot enough for continual H$_2$ thermal desorption and efficient photothermal decomposition at a rate of 3-5 photons per H$_2$ molecule. Here, the photocatalyst exhibits linear behavior, and delaying the second pulse provides little to no enhancement over the undelayed case. By varying the partial pressure, we find that dissociative adsorption of NH$_3$ is usually the RDS, and by varying the relative pulse intensities, we find that a stronger second pulse is favored by the nonthermal process that helps desorb H$_2$. 

This P3C technique, which requires no special tools beyond a motorized translation stage and a mirror with variable reflectivity, may be applied universally to any photocatalytic reaction using any photocatalyst.  The ability to vary excitation intensity, wavelength, external temperature, reactant pressure, and pulse asymmetry as a function of pulse delay provides unprecedented insight into the underlying mechanisms and affords a unique opportunity to optimize the distribution of incident photons to maximize desired products.

\section{Methods}\label{sec11}
\par \underline{Catalyst Preparation}
\par\noindent Cu-Ru surface alloy supported on MgO-Al$_2$O$_3$(Cu-Ru-AR, 19.5 at\% Cu\& 0.5 at\% Ru): The Cu-Ru surface alloy supported on MgO-Al$_2$O$_3$ was synthesized using a co-precipitation method.  The metal precursor was synthesized dissolving  0.707 g (2.925 mmol) Cu(NO$_3$)$_2$·3H$_2$O (Sigma-Aldrich, \#61194), 0.0190 g (0.075 mmol) RuCl$_3$·xH$_2$O (Sigma-Aldrich, \#206229), 2.308 g (9 mmol) Mg(NO$_3$)$_2$·6H$_2$O (Sigma-Aldrich, \#63084) and 1.125 g (3 mmol) Al(NO$_3$)$_2$·9H$_2$O (Sigma-Aldrich, \#237973) in 15 mL DI water. The basic solution was prepared by dissolving 2.544 g (24 mmol) anhydrous Na$_2$CO$_3$ in 20 mL DI water.  The metal precursor solution and Na$_2$CO$_3$ solution were added drop wise simultaneously to 100 mL of 80$^\circ$C DI water in a 5-neck, round bottom flask while maintaining  pH of 8 (Accument, AP63). In order to maintain the pH of 8, the addition of metal precursor and Na$_2$CO$_3$ solution were varied over the course of 20 minutes. The resulting pH=8 solution was left to stir for 24 hrs at  80$^\circ$C before cooling to room temperature. The Cu-Ru catalyst was isolated via centrifugation (1000 G), washing with Milli Q water 5 times and subsequently drying under air at 120$^\circ$C  overnight. In order to ensure purity, the catalyst was annealed at 600$^\circ$C under a constant flow of 50 sccm He (Airgas, ultrahigh purity, 99.999\%) and subsequently reduced at 500$^\circ$C under a constant flow of 20 sccm H$_2$ (Airgas, research purity, 99.9999\%) for 1 hr. 
\bigskip
\par \noindent \underline{Catalyst Characterization}
\par \noindent \underline{High-angle annual dark-field scanning transmission electron microscopy (HAADF-STEM)} High angle annular dark field scattering transmission electron microscopy (HAADF-STEM) was performed with the FEI Titan Themis scanning transmission electron microscope with spherical and chromatic aberration correctors. The sample was pretreated by annealing at 500$^\circ$C in He followed by a reduction at 500$^\circ$C in H$_2$ prior to suspending the sample in Milli-Q water and drop casting on a nickel lacey carbon TEM grid. The Cu$_{19.5}$Ru$_{0.5}$ nanoparticles within the catalyst are spherical in morphology (Fig. S1A) with an average size of 7.6 nm (Fig. S1B). Elemental mapping (EDX) indicates the presence of alumina, magnesia, and oxygen which are present within the support, and copper nanoparticles (Fig. S2). The ruthenium exists as only 0.5\% of the sample which is why there is limited signal for this element in the EDX. High resolution HAADF-STEM was used to observe ruthenium clusters and single atoms which appear brighter in comparison to copper due to differences in electron density (Fig. S3). Additionally, the d-spacing within the lattice was measured using ImageJ software and confirmed the presence of (111) face centered cubic crystalline copper (PDF-04-0836) which corresponds well with the XRD data in (Fig. S5). 
\par \noindent \underline{Diffuse Reflectance} was measured in an Agilent Cary 5000 UV-Vis-NIR spectrometer. A Praying Mantis diffuse-reflection accessory (Harrick Scientific Products Inc., DRP-VA) was attached to the spectrometer to convert from transmission mode to diffuse-reflection mode. After annealing and reduction, the sample was quickly transferred to the Praying Mantis accessory. MgO powder was used as a reference for collecting the background spectrum (Figure S3).  
\par \noindent \underline{Power X-ray diffraction (PXRD) spectra} were measured on a Rigaku SmartLab X-ray Diffractometer instrument. Spectra were collected from 30$^\circ$ to 100$^\circ$ at a speed of 2$^\circ$/min. Samples were reduced before being quickly transferred onto the PXRD sample holder.
\par \noindent \underline{X-ray Photoelectron Spectroscopy (XPS)} was performed on a Thermo Scientific Nexsa G2 X-Ray Photoelectron Spectrometer. Catalysts were reduced and quickly transferred onto carbon tape. The C 1s (284.4 eV) level was used to calibrate the system work function. XPS was used to study the chemical environment of Cu and Ru. 

\bigskip
\par \noindent \underline{NH$_3$ Decomposition Experiments}
\par \noindent \underline{Pump-Pump Photocatalysis} was performed in the Harrick reaction chamber (High Temperature raman cell), which allows for laser illumination through the CaF$_2$ optical window. A Ti:Sapphire laser (Coherent Chameleon) paired with a second harmonic generation system (Coherent VUE SHG) was used as the illumination source for all experiments. The pump-pump delay was achieved by splitting the laser with a 45/55 pellicle (ThorLabs, \#BP145B1) to an electronic delay stage (ThorLabs, \#LTS300C) (Figure S6). The reactant gas NH$_3$ (Airgas, anhydrous purity, 99.99\%), as well as the preparation gases, were fed into the reaction chamber through flow rate controllers (Allicat Scientific, Inc.). The effluent gas was analyzed by gas chromatography (Shimadazu, Inc., GC-2014) with a PDHID detector. 
\par \noindent \underline{Time Zero} was identified using a broadband reverse saturable absorbing (RSA) organometallic complex within the range of 400-800 nm. 
\par \noindent \underline{Partial Pressure dependence studies} were completed by adjusting the NH$_3$ partial pressures through the addition of He and/or Ar gas. Due to the varying thermal conductivity of these gases, a ratio of He/Ar was used to maintain the same surface temperature as pure NH$_3$. Illumination was constant throughout the experiment and the products were analyzed via gas chromatography. The NH$_3$ partial pressure changed from 0.2 to 1 atm while the power was varied for photocatalysis and the temperature for thermocatalysis. 
\par \noindent \underline{Power-dependence studies} were completed utilizing a neutral density filter altering the power from 40 mW to 120 mW. This experimental power was reduced by 12.5\% when the reflections from the CaF$_2$ window, MgO, and Al$_2$O$_3$ were taken into account.

\backmatter

\bmhead{Supplementary information}
Supplemental Figures S1-S8

\bmhead{Acknowledgements}
 The authors thank Catherine Arndt and William Smith for their early contributions to P3C. 
\section*{Declarations}
\paragraph*{Funding:}
C.J.F. acknowledges support from the Department of Defense SMART Scholarship under OUSD/R\&E, NDEP/BA-1. This work was supported by the Air Force Office of Scientific Research under award number FA9550-42-1-0178 (N.J.H. and P.N.) and the Welch Foundation C-1220 (N.J.H.) and C-1222 (P.N.).
\paragraph*{Author contributions:}
HOE conceived P3C and guided the experiments with NJH.  CJF constructed the experimental apparatus, performed the photocatalytic measurements, and created the figures.  SD performed the TEM imaging, YY fabricated the catalysts, and LK performed XPS and XRD measurements. All authors contributed to the writing of the manuscript.
\paragraph*{Competing interests:}
There are no competing interests to declare.






\bibliography{sn-bibliography}

\end{document}